\documentclass{PoS}

\usepackage{graphicx}
\usepackage[percent]{overpic}
\usepackage{caption}
\usepackage{gensymb}
\usepackage{subcaption}

\newcommand{\apj}{ApJ}

\newcommand{\apjl}{ApJL}

\newcommand{\Fermi}{\emph{Fermi}}  
\newcommand{\FermiLat}{\emph{Fermi} LAT}     

\newcommand{\g}{$\gamma$}
\newcommand{\st}{$^{\mathrm{st}}$}  
\newcommand{\nassocprobclassified}{$36$}
\newcommand{\nclassifiedsnrs}{$30$} 

\newcommand{\nmarginalAIEM}{two} 
\newcommand{\nGalSNRs}{$279$} 
\newcommand{\nnotsnrs}{four} 
\newcommand{\ndetected}{$102$} 
\newcommand{\nULs}{$245$} 
\newcommand{\nextended}{$17$} 
\newcommand{\nnewpointlike}{$10$} 
\newcommand{\nnewextended}{four} 

\title{The First \FermiLat~ SNR Catalog SNR and Cosmic Ray Implications}

\ShortTitle{The First Fermi-LAT SNR Catalog SNR and Cosmic Ray Implications}

\author{
T.~J.~Brandt$^{\;(1)\;}$, 
F.~Acero$^{\;(2)\;}$, 
\speaker{F.~de Palma}$^{\;\; \dagger \,(3,4)\;}$, 
J.~W.~Hewitt$^{\;(1,5)\;}$, 
M.~Renaud$^{\;(6)\;}$,$\quad$ $\quad$ $\quad$ $\quad$ $\quad$ $\quad$
on behalf of the \FermiLat~ Collaboration \\
$^{(1)}$ NASA Goddard Space Flight Center, Greenbelt, MD 20771, USA \\
$^{(2)}$ Laboratoire AIM, CEA-IRFU/CNRS/Universit\'e Paris Diderot, Service dÕAstrophysique, CEA Saclay, 91191 Gif sur Yvette, France \\
$^{(3)}$ Istituto Nazionale di Fisica Nucleare, Sezione di Bari, 
Bari, Italy \\
$^{(4)}$ Universit\'a Telematica Pegaso, Piazza Trieste e Trento, 48, 80132 Napoli, Italy \\
$^{(5)}$ CRESST/University of Maryland, Baltimore County, Baltimore, MD 21250, USA \\
$^{(6)}$ Laboratoire Univers et Particules de Montpellier, Universit\`e Montpellier 2, CNRS/IN2P3, Montpellier, France \\
$\dagger$ {\footnotesize{E-mail:}} {\tt{\footnotesize{francesco.depalma@ba.infn.it}}}}

\abstract{Galactic cosmic ray (CRs) sources, classically proposed to be Supernova Remnants (SNRs), must meet the energetic particle content required by direct measurements of high energy CRs. Indirect gamma-ray measurements of SNRs with the \Fermi~Large Area Telescope (LAT) have now shown directly that at least three SNRs accelerate protons. With the first Fermi LAT SNR Catalog, we have systematically characterized the GeV gamma-rays emitted by 279 SNRs known primarily from radio surveys. We present these sources in a multiwavelength context, including studies of correlations between GeV and radio size, flux, and index, TeV index, and age and environment tracers, in order to better understand effects of evolution and environment on the GeV emission. We show that previously sufficient models of SNRs' GeV emission no longer adequately describe the data. To address the question of CR origins, we also examine the SNRs' maximal CR contribution assuming the GeV emission arises solely from proton interactions. Improved breadth and quality of multiwavelength data, including distances and local densities, and more, higher resolution gamma-ray data with correspondingly improved Galactic diffuse models will strengthen this constraint. }

\FullConference{The 34th International Cosmic Ray Conference,\\
		30 July- 6 August, 2015\\
		The Hague, The Netherlands}

\begin{document}

\section{Introduction}
In this work we will shortly describe the results obtained from the systematic analysis of supernova remnants (SNR) in 3 years of \Fermi-Large Area Telescope (LAT) data. We will correlate our spectral and spatial results with those obtained in other wavelength, trying to better understand SNRs' emitting mechanisms, ages and environments. At the end of this work we will discuss under some assumptions the contribution of the SNRs to the observed cosmic rays (CR). The method of this analysis is described in \cite{ICRC_poster} while the full analysis will be shown in a future publication \cite{snrcat}. 
\section{Summary of results}
\begin{figure}[h]
\centering
\begin{overpic}[width=0.5\columnwidth]{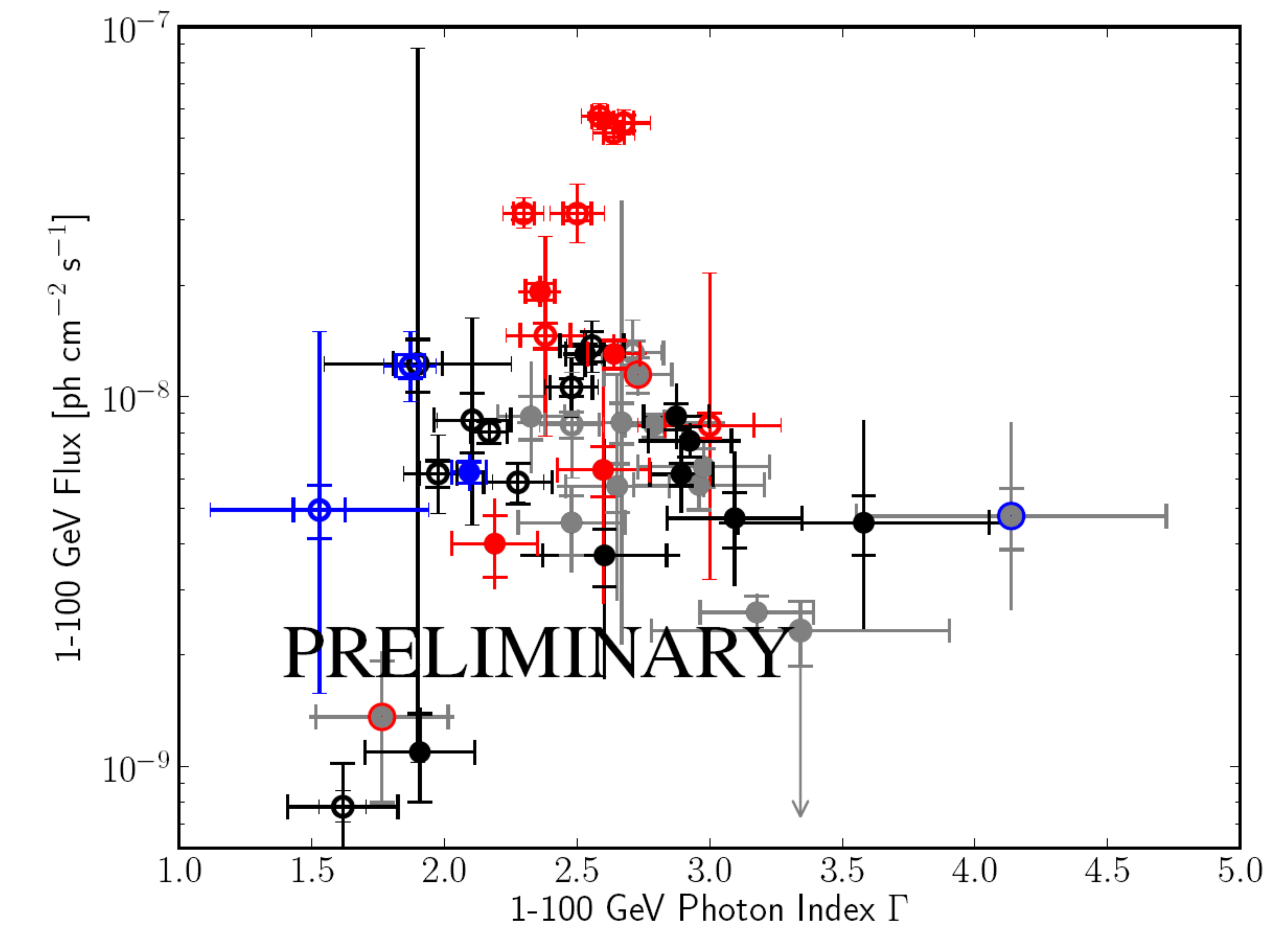}
\end{overpic}
\caption{ The distribution of fitted photon index and flux in the energy range $1-100$\,GeV. 
Open circles indicate extended SNRs while filled circles indicate point-like sources. All SNRs that passed classification are shown as black unless also classified as young non-thermal X-ray SNRs (blue) or as interacting with MCs (red). Candidates which did not pass classification but still had both fractional overlaps $>0.1$ are grey. If they are also young or interacting, they are outlined in blue or red, respectively. 
Statistical error bars have caps; error bars without caps present the systematic error, described in \cite{ICRC_poster}. The same color code is used in all the following plots.
\label{fig:GeVFluxGeVIndex}}
\end{figure}

Using three years of data and starting from the spatial information of \nGalSNRs~SNRs \cite{greencat} we detected \ndetected~candidates with a final source TS $>25$. Of the \ndetected~detected candidates, \nassocprobclassified{} passed our association probability threshold. Of these, \nclassifiedsnrs~SNRs ($\sim11\%$ of the total) show significant emission for all alternative interstellar emission models (IEM, see also \cite{dePalma13-AltIEMSystematics_FSymp}) and are classified as likely GeV SNRs. An additional \nnotsnrs{} were identified as sources which are not SNRs; \nmarginalAIEM~other candidates were demoted to marginal due to their dependence on the IEM.
Of the sources likely to be GeV SNRs, \nextended~show evidence for extension (TS$_{ext} > 16$). Only sources associated with SNRs G34.7$-$0.4 and G189.1+3.0 show evidence of significant spectral curvature in the $1-100$\,GeV range and are fit with logParabola spectra. 
Of the classified candidates, \nnewextended~extended and \nnewpointlike~point SNRs are new. For all the sources we evaluated their spectral and spatial characteristics with systematic and statistical uncertainties. We reported \nULs ~Bayesian \cite{Helene83} upper limits (UL) for all the non-detected SNRs. In Figure~\ref{fig:GeVFluxGeVIndex} the observed spectral parameters of all the SNRs are shown: the photon flux range is larger than two order of magnitude while the power-law (PL) index range from $1.5$ to $\sim 4.0$.

\section{Multiwavelength correlations}
\begin{figure}
 \centering
 \begin{overpic}[width=0.5\columnwidth]{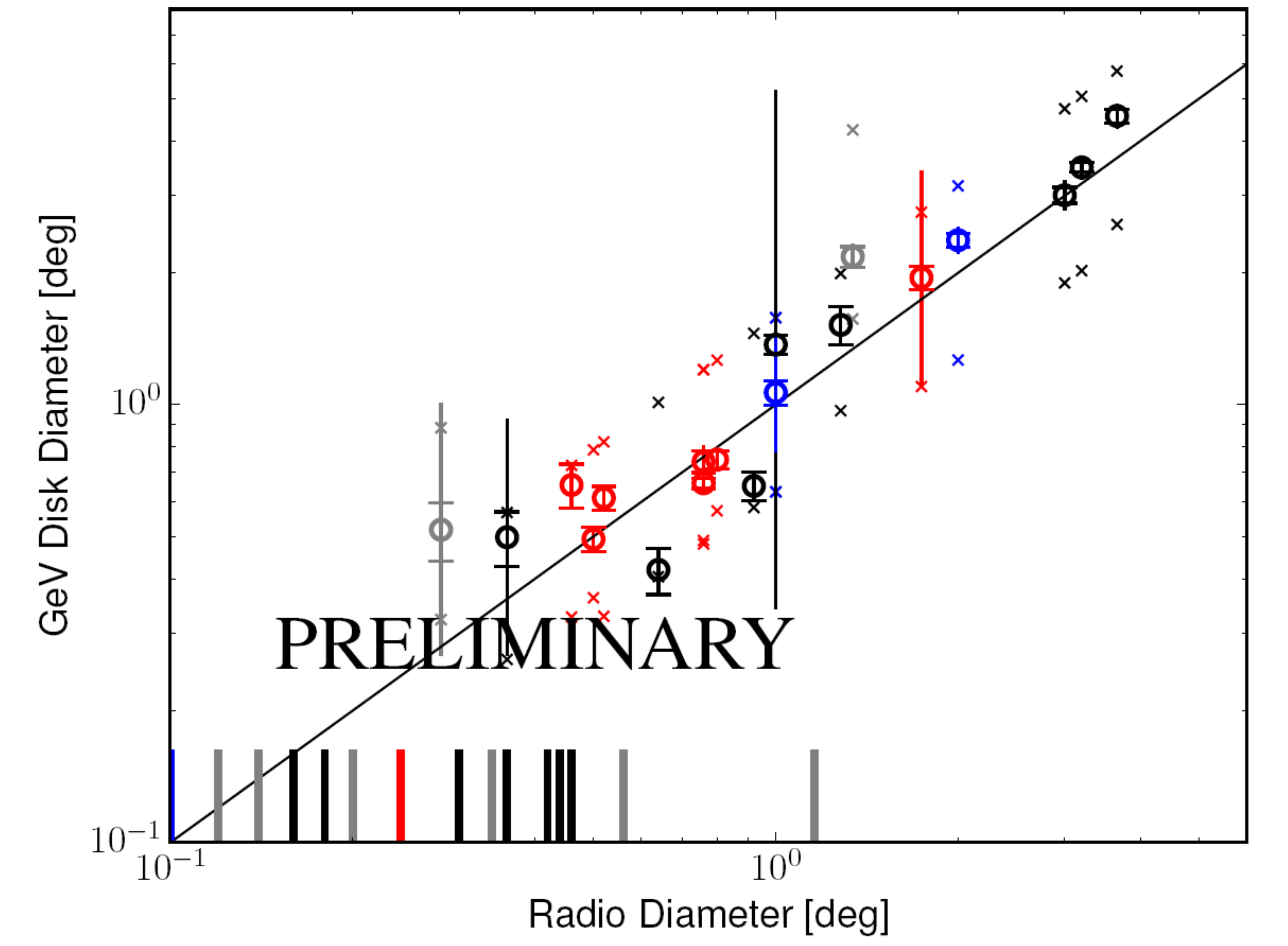}
 \end{overpic}
 \caption{The radio diameters of the SNRs from Green's catalog are correlated with the fitted GeV diameters for those candidates with significant extension. The solid line represents equal radio and GeV diameters. All cases of detected extension have diameters greater than $0.2$\degree. The ticks denote the radio extension of GeV point-like candidates, colored in order of their characteristics (young or interacting) and by their classifications (well defined or marginal). The small `x's bracketing the points show the minimum and maximum GeV extensions allowed such that the source remains classified or marginally classified given the radio position and extension and best fit GeV position, see \cite{snrcat}.}
 \label{fig:GeVradioSize}
 \end{figure}

 We find that the best-fit GeV diameter is within errors of the radio diameter for most of the candidates classified as associated with an SNR, as shown in Figure~\ref{fig:GeVradioSize}. This result support the hypotheses which have both GeV and radio emission arising from the same location, e.g. \cite{Uchiyama10-crushedClouds}. 
 
\begin{figure}[ht]
  \centering
  \begin{subfigure}{.49\textwidth}
  \centering
  \begin{overpic}[width=\linewidth]{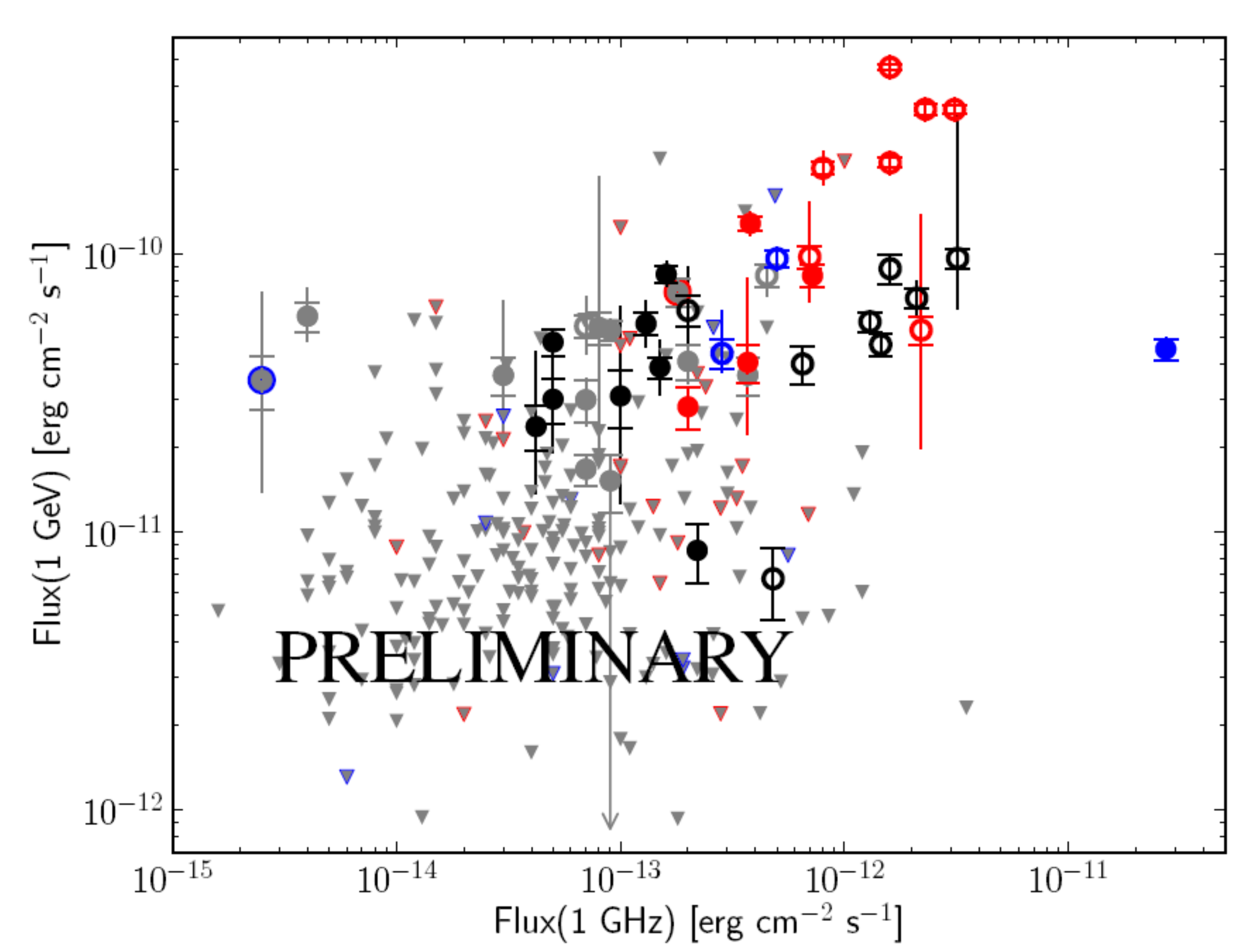}
  
  \end{overpic}
  \subcaption{Flux comparison. }
  \label{fig:GeVradioFlux}
  \end{subfigure}
  \begin{subfigure}{.475\textwidth}
  \begin{overpic}[width=\linewidth]{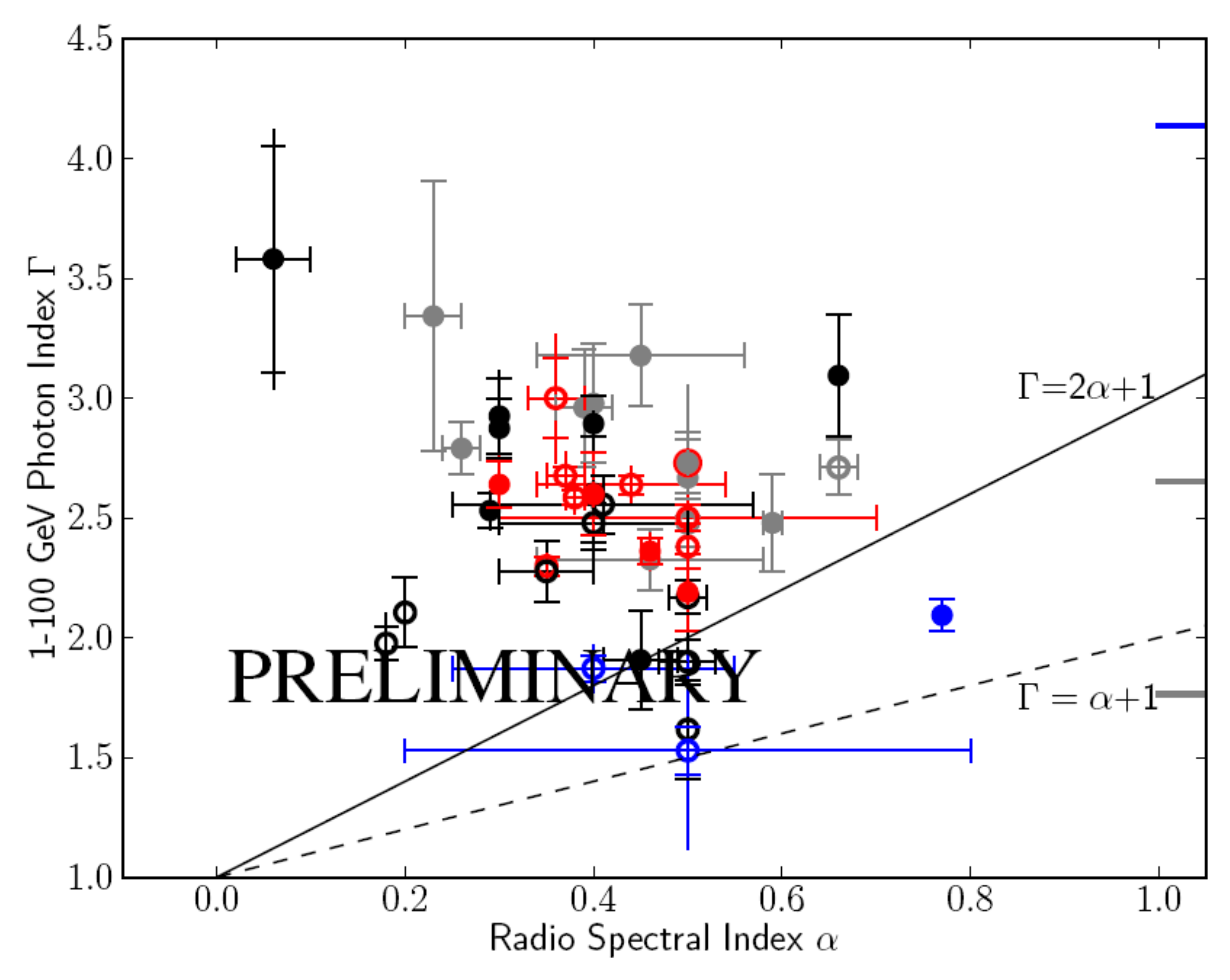}  
  \end{overpic}
  \subcaption{Index comparison. }
  \label{fig:GeVradioIndex}
\end{subfigure}
   \caption{{\it Left:} comparison of \g-ray and radio spectral flux densities for all SNRs and candidates. For all SNRs that were not detected or which failed classification, grey triangles indicate upper limits at $99\%$ confidence, computed assuming the radio location and extension. {\it Right:} comparison of radio spectal index, $\alpha$, and GeV photon index, $\Gamma$. The expected correlations for $\pi^0$ decay or e$^{\pm}$ bremsstrahlung (solid) and inverse Compton (dashed) are plotted. Ticks along the right hand side of the right plot show the $1-100$\,GeV photon indices of those SNRs without reported radio spectral indices. }
\end{figure}

 Figure~\ref{fig:GeVradioFlux} shows the flux from synchrotron radio emission at $1$\,GHz in comparison to the \g-ray flux at $1$\,GeV.
 A correlation may exist between the radio and GeV flux \cite{Uchiyama10-crushedClouds}, if the same lepton population is directly producing both emissions or if they scale with some underlying physical parameter such as ambient density.  In \cite{snrcat} we do not find evidence for a significant correlation, but we also cannot strictly rule out an intrinsic correlation for various effects (e.g. small energy range used for the GeV analysis, possible change in the spectral index etc).

 Figure \ref{fig:GeVradioIndex} compares the deduced radio spectral index $\alpha$ with the $1-100$\,GeV photon index $\Gamma$, theoretical predictions for two emitting mechanism are shown on the plot (see label). Nearly all candidates have softer-than-predicted \g-ray photon indices given their radio spectra. The three young SNRs in blue are most consistent with a single underlying particle population, and it has been suggested they emit via inverse compton (IC, dashed line) at GeV energies. The observed soft GeV spectra relative to the radio has several potential explanations (e.g. different PL index for leptonic and hadronic populations, the emitting particle spetra has breaks or different zones with different properties dominating the emission at different wavelengths).

\begin{figure}[h]
  \centering
   \begin{overpic}[width=0.5\columnwidth]{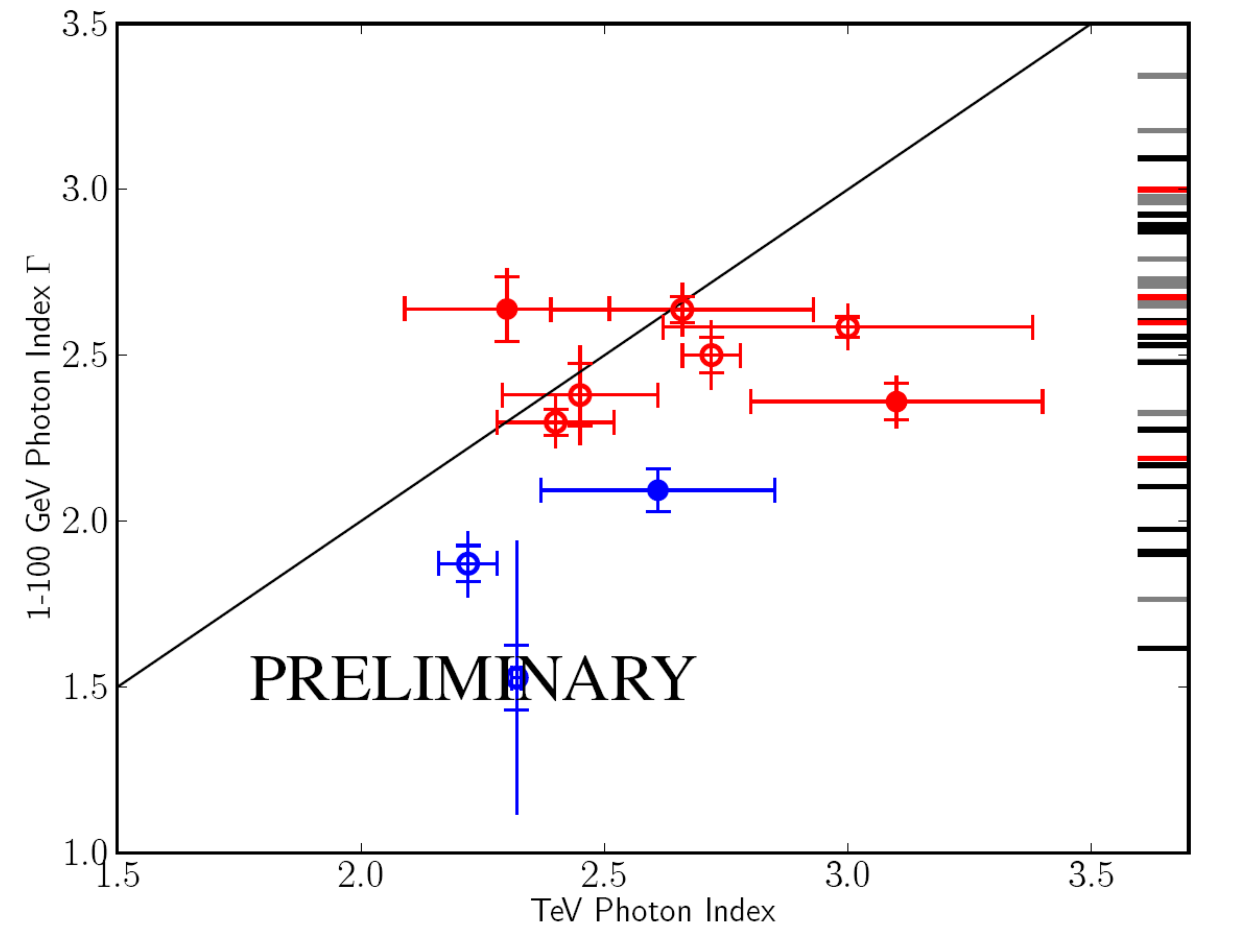}
 \end{overpic} 
  \caption{GeV index compared to published index measurements from Imaging Atmospheric  Cherenkov Telescopes (IACT). The line corresponds to equal index values. The ticks represent the GeV candidates with indices in the range of those with a TeV counterpart but with no TeV measurements themselves.  Yet many of these are hard: $12$ candidates and $10$ marginal candidates have indices harder than $2.5$, suggesting they may well be observable a TeV energies.}
  \label{fig:GeVTeVIndex}
 \end{figure}

 In Figure~\ref{fig:GeVTeVIndex} we plot the PL index in the GeV versus TeV range for all SNRs observed, the majority of them seems to have TeV indices that are softer than their GeV indices. This characteristic suggests spectral curvature, potentially reflecting a change in spectral slope of the underlying particle population(s') index or indices, or that another particle population dominates at higher energies or the emssion mecchanism is different.
 \section{Age Vs. Environment}
 
\begin{figure}[ht]
  \centering
  \begin{subfigure}{.475\textwidth}
  \centering
  \begin{overpic}[width=\linewidth]{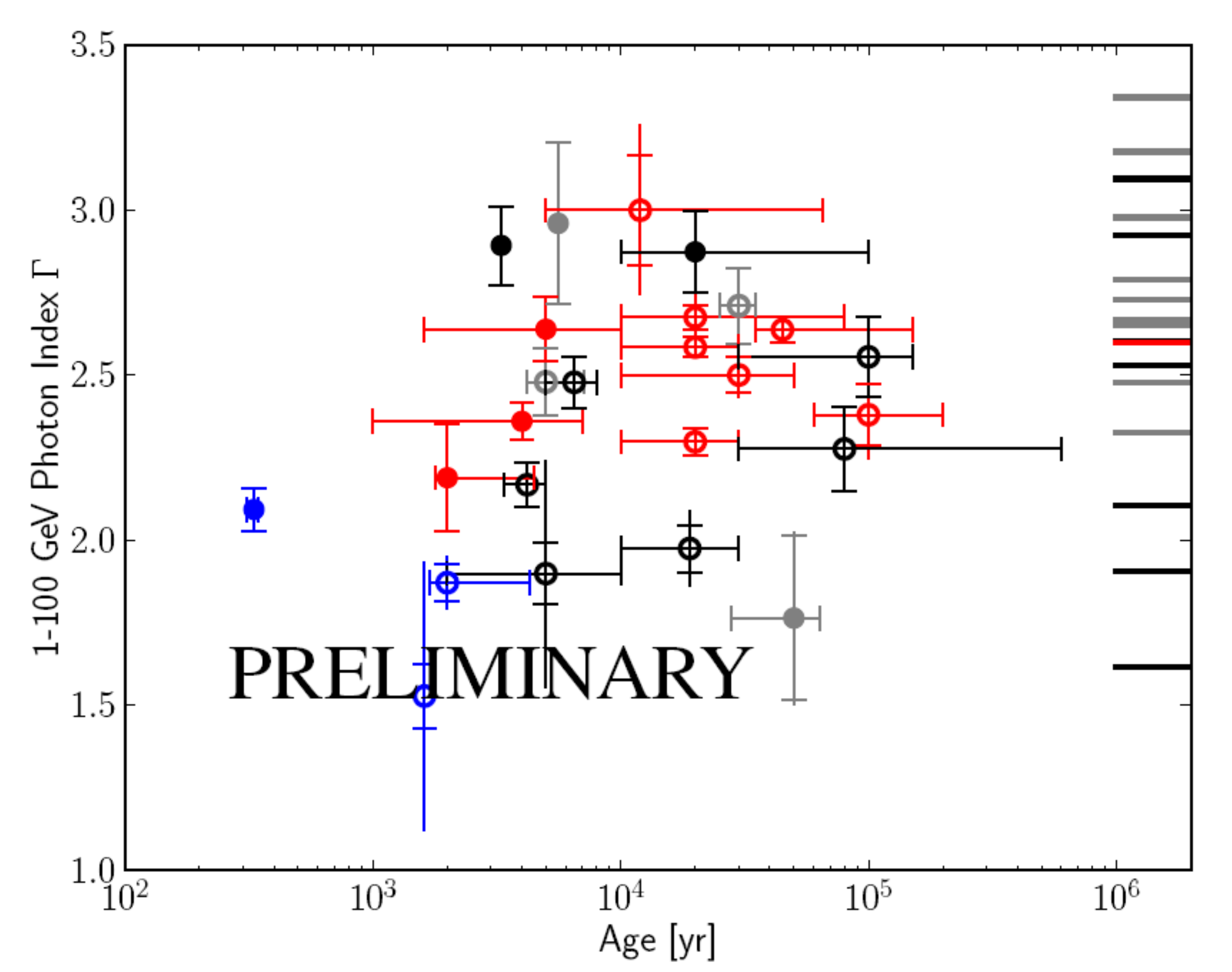}
  
  \end{overpic}
  \subcaption{Age versus GeV spectral index \label{fig:AgeGeVIndex}}
  \end{subfigure}
  \begin{subfigure}{.49\textwidth}
  \begin{overpic}[width=\linewidth]{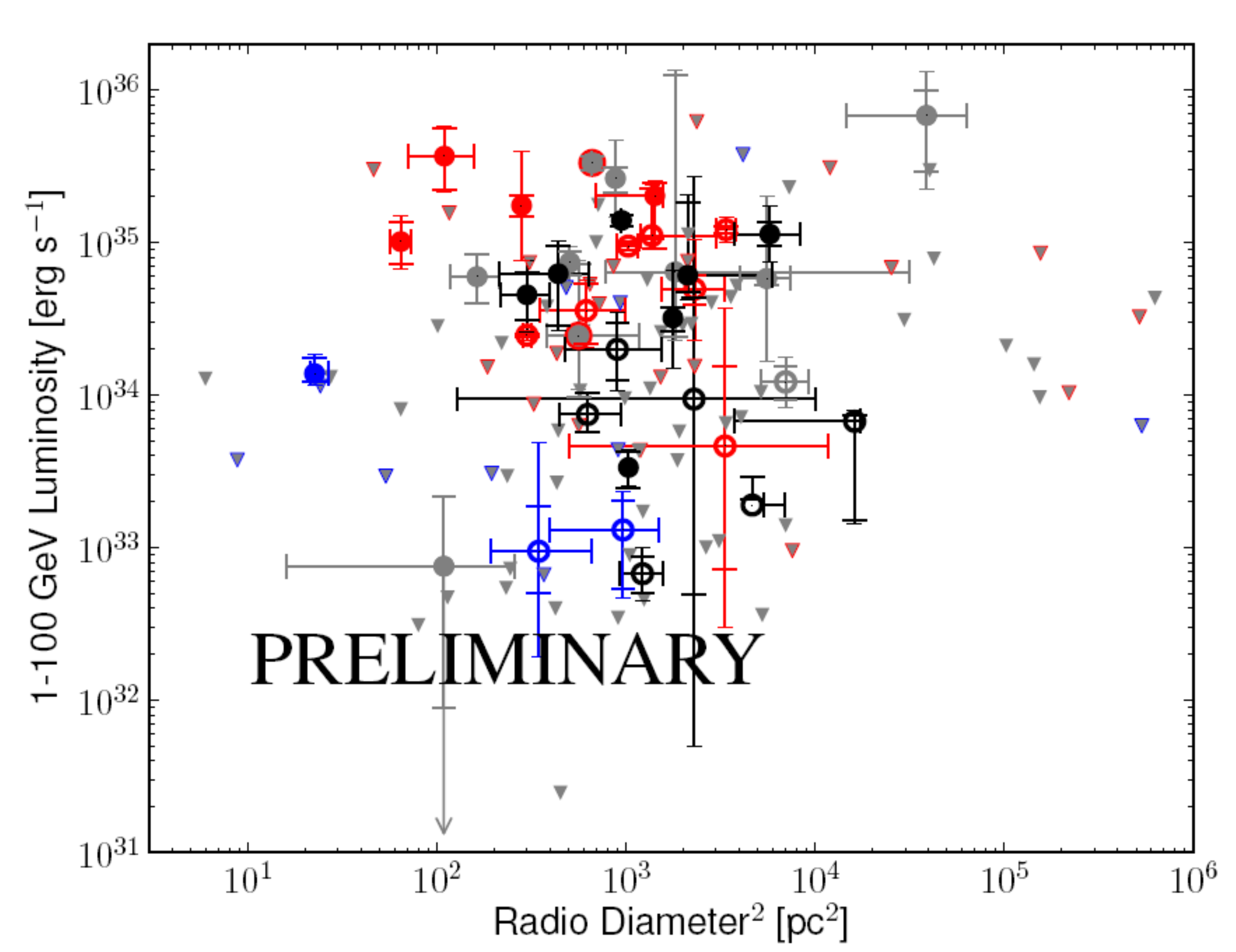}  
  \end{overpic}
    \subcaption{GeV luminosity versus  $(\mathrm{radio~diameters})^2$.\label{fig:LumDia}}
\end{subfigure}
   \caption{{\it Left:} Age versus GeV spectral index. For those with ages in the literature, the young (blue) SNR candidates are separated in this phase space from the identified interacting candidates (red). The ticks on the right show indices for GeV candidates without well-established ages. {\it Right:} The $1-100$\,GeV luminosity is plotted against the square of the radio diameters in pc of those SNRs with known distances.}
\end{figure}

In Figure~\ref{fig:AgeGeVIndex} the SNR ages from literature versus the $1-100$\ GeV photon indexes are shown. Young SNRs tend to have harder GeV photon indices than interacting SNRs, which are likely middle aged, though the scatter in age for the two classes is one to two orders of magnitude. This trend may be due to the decrease of the maximum acceleration energy as SNRs age and their shock speeds slow down or the interacting SNRs may be more luminous due to their interactions with denser surroundings not yet reached by younger SNRs. Ultimately, the large scatter observed in luminosity will likely reflect effects due to both age and environment.

To investigate the role of environment in the trends for the young and interacting SNRs, we examined the GeV luminosity versus radio diameter in Figure~\ref{fig:LumDia}. The square of the physical diameter ($D$) can be regarded as a reasonable indicator for SNR age and environment, see \cite{snrcat}. We note that any apparent correlation between the luminosity and $D^2$ may be due to their inherent dependence on distance (squared). Figure~\ref{fig:LumDia} shows that, for the detected candidates, interacting SNRs are generally more luminous for a given physical diameter than young SNRs, though there is large scatter. This suggests that SNRs at the same age may be more luminous because they have encountered denser gas. With the addition of upper limits, we find some interacting candidates are constrained to lie below the luminosities of most young SNRs. Thus, as we continue to detect SNRs with increasingly fainter \g-ray fluxes, we are likely to find less separation between the luminosities of the two classes.

\section{Constraining Cosmic Ray acceleration}

 This systematic search for \g-ray emission from known SNRs provides an opportunity to understand if Galactic SNR population is able to accelerate and release CRs with the appropriate composition and flux up to the transition between the Galactic and extragalactic components.
 In the following, we assume that the \g-ray emission from SNRs probed with LAT entirely arises from the interaction of CR protons and nuclei with the surrounding ISM or circumstellar medium through the production and subsequent decay of $\pi^0$. Given that two other emission mechanisms involving accelerated leptons, namely non-thermal bremsstrahlung and IC scattering, could also contribute in the \g-ray domain, the constraints derived from the LAT measurements should be considered as upper limits on the CR energy content in SNRs.

\begin{figure}[]
  \centering
  \begin{subfigure}{.49\textwidth}
  \centering
  \begin{overpic}[width=\linewidth]{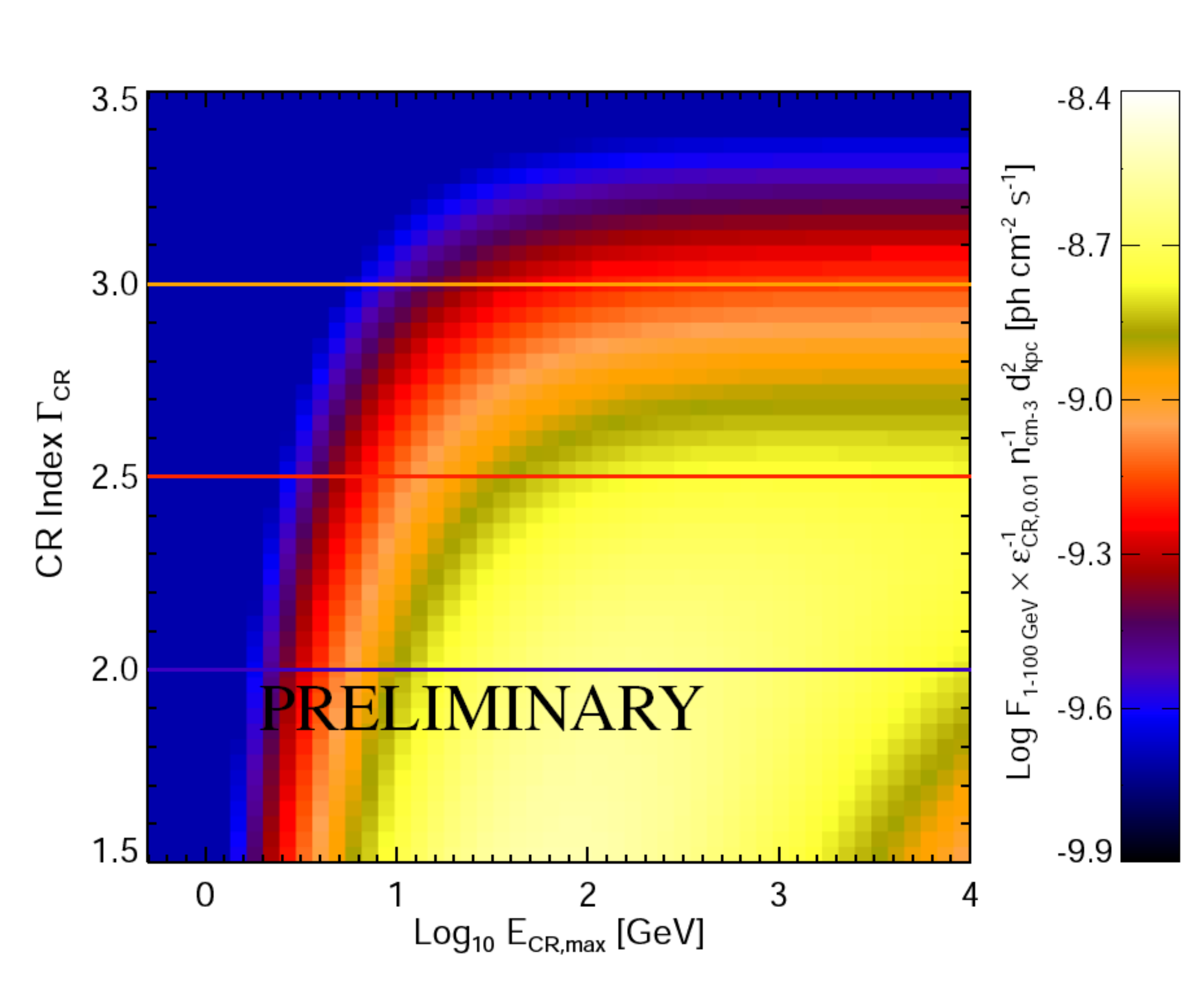}
  
  \end{overpic}
  \subcaption{\label{fig:CR_FluxvsEmaxAndIndex_1}}
  \end{subfigure}
  \begin{subfigure}{.475\textwidth}
  \begin{overpic}[width=\linewidth]{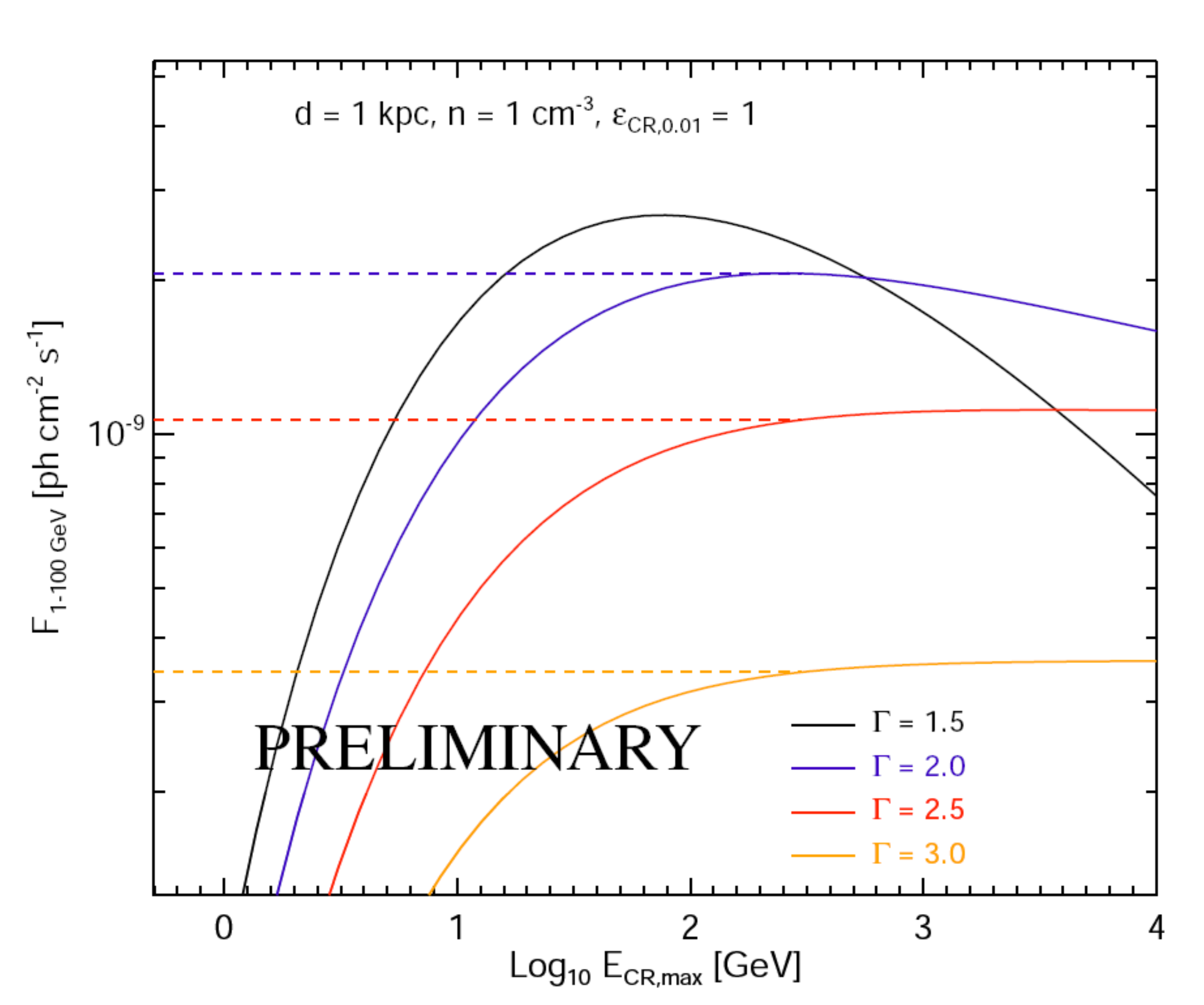}  
  \end{overpic}
  \subcaption{\label{fig:CR_FluxvsEmaxAndIndex_2}}
  
\end{subfigure}
   \caption{{\it Left:} Under standard assumptions (see text), an SNR's \g-ray flux in the $1-100$\,GeV range can be related to the accelerated CRs' maximal energy $E_{\mathrm{CR, max}}$ and spectral index $\Gamma_{\mathrm{CR}}$ for a given CR energy content above a particle momenta of $10$\,MeV\,$c^{-1}$ ($\epsilon_{\mathrm{CR}} = E_{\mathrm{CR}}/E_{\mathrm{SN}} = 0.01$), effective density ($1$\,cm$^{-3}$), and distance to the SNR ($1$\,kpc). {\it Right:} the relationship between the SNR's \g-ray flux in the $1-100$\,GeV band and $E_{\mathrm{CR, max}}$ for different values of $\Gamma_{\mathrm{CR}}$. For $E_{\mathrm{CR, max}}\gtrsim 200$\,GeV and $\Gamma_{\mathrm{CR}}\gtrsim 2$, the flux is weakly dependent on the CR maximal energy.}
\label{fig:CR_FluxvsEmaxAndIndex}
\end{figure}

Figure~\ref{fig:CR_FluxvsEmaxAndIndex_1} shows the \g-ray flux from p-p interactions \cite{Kamae06-PPinteractions} in the ($E_{\mathrm{CR, max}}$,~$\Gamma_{\mathrm{CR}}$) plane and shows its dependence on $E_{\mathrm{CR, max}}$ for different CR spectral indices, with $d =1$\,kpc, $n =1$\,cm$^{-3}$, $\epsilon_{\mathrm{CR}} =0.01$, and $E_{\mathrm{SN}}=10^{51}$\,erg (for further details see \cite{snrcat}). The CR energy content is computed for particle momenta above $10$\,MeV\,$c^{-1}$ and is related to the total supernova explosion energy with the relation $E_{\mathrm{CR}} \equiv \epsilon_{\mathrm{CR}}E_{\mathrm{SN}}$. The \g-ray flux is nearly independent of the CR maximal energy as long as $E_{\mathrm{CR, max}}\gtrsim 200$\,GeV and $\Gamma_{\mathrm{CR}}\gtrsim 2$. In this case, the \g-ray flux can conveniently be approximated and parametrized as shown in Figure~\ref{fig:CR_FluxvsEmaxAndIndex_2}. Details on this parametrization and references can be found in \cite{snrcat}. We use the relationship between an SNR's \g-ray flux, density, and distance shown in Figure~\ref{fig:CR_FluxvsEmaxAndIndex_2} to determine the maximal CR energy content $E_{\mathrm{CR}}$ through $\epsilon_{\mathrm{CR}}$ contributed by every SNR for which we have measured the \g-ray flux and photon index or derived an upper limit at the $95$\% confidence level. In the case of a detected SNR, the photon index, known to reproduce the spectral shape of the parent CR proton/nuclei spectrum above a photon energy of $1$\,GeV \cite{Kamae06-PPinteractions}, is taken to be equal to $\Gamma_{\mathrm{CR}}$. In the case of the upper limits, we assume an index of $2.5$ (i.e.~the average value of the detected SNRs). We use the canonical value of $10^{51}$\,erg for $E_{\mathrm{SN}}$. 

\begin{figure}[]
\centering
\begin{overpic}[width=0.8\columnwidth]{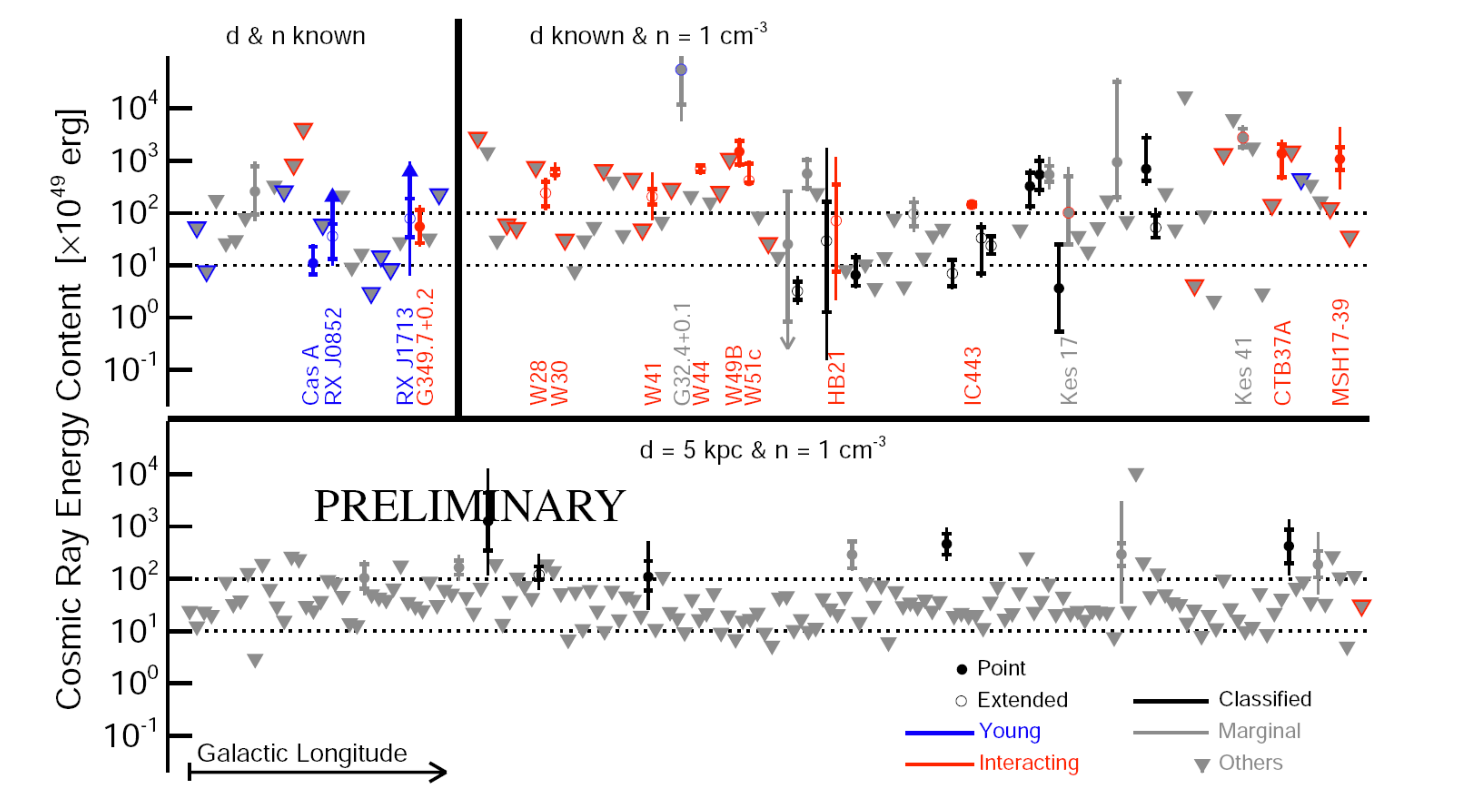}
  \end{overpic}
\caption{Estimates of the CR energy content (in units of $10^{49}$\,erg) for all Galactic SNRs, divided into three categories each sorted by Galactic longitude: SNRs with existing distance and density estimates (upper left panel); SNRs with known distances (upper right panel); and SNRs with unknown distance and density (lower panel). The two dashed lines indicate a CR energy content of $10$ and $100\%$ of the standard SN explosion energy.}
\label{fig:CREfficiencies}
\end{figure}

Translating SNRs' \g-ray measurements into constraints on their contribution to $E_{\mathrm{CR}}$ also requires knowledge of their distances and effective densities, both quantities are taken from the literature. Figure~\ref{fig:CREfficiencies} shows the constraints on the CR energy content for the population of known Galactic SNRs. As is clearly visible for the first two subclasses of SNRs, the estimates and upper limits on the CR energy content span more than three orders of magnitude, from a few\,$\times10^{49}$\,erg to several\,$\times10^{52}$\,erg.  For the interacting SNRs that lie above the $\epsilon_{\mathrm{CR}} = 1$ dashed line, the densities experienced by the CR particles in the molecular cloud (MC) interaction region are likely much larger than those used in this analsysis. Interacting candidates' lying above this limit suggests that they are likely the sites of hadronic interactions in dense environments. In contrast, most of the young SNRs lie at or below this luminosity limit, suggesting that IC processes may contribute to their measured luminosity. New information about distances and densities can give us further insight on the ability of known SNRs to provide the observed CRs.
The usual assumption of $\epsilon_{\mathrm{CR}} = 0.1$, required in order for the Galactic SNR population to supply the CR flux observed at Earth, is compatible with the results of the \FermiLat SNR catalog, for more details see \cite{snrcat}.

\section{Conclusion}
We examined our GeV candidate SNR population in light of multiwavelength observations in order to better understand both SNRs' characteristics and potential for accelerating hadrons. In this work we have studied correlation of our results with spectral and spatial information from radio and TeV bands, in order to understand if the underlying CR population is the same and has the same spectral characteristics. Within the limits of existing MW data, our observations generally support previous findings of changes in spectral slope at or near TeV energies and a softening and brightening in the GeV range with age, yet we see indications that new candidates and new multiwavelength data may provide evidence of exceptions to this trend. In examining tracers of age and environment, we found that the classified candidates followed the previously observed trends of young SNRs being harder and fainter at GeV energies than older, often interacting SNRs.

We find, also, that the limits on CR energy content span more than three decades, including many interacting candidates for which the densities in the interaction regions are much greater than the nominal density assumed in the calculation, and young candidates with efficiencies below the nominal $\sim10\%$, consistent with possible leptonic emission predictions (e.g. IC). The contribution from all SNRs, particularly those with flux upper limits, is beginning to constrain the energy content put into CRs from the known SNRs to less than $10\%$, particularly in regions of well characterized IEM background. We find that the candidates and upper limits are generally within expectations if SNRs provide the majority of Galactic CRs.

\section{Acknowledgments}

The \textit{Fermi}-LAT Collaboration acknowledges support for LAT development, operation and data analysis from NASA and DOE (United States), CEA/Irfu and IN2P3/CNRS (France), ASI and INFN (Italy), MEXT, KEK, and JAXA (Japan), and the K.A.~Wallenberg Foundation, the Swedish Research Council and the National Space Board (Sweden). Science analysis support in the operations phase from INAF (Italy) and CNES (France) is also gratefully acknowledged.

\end{document}